\documentclass{emulateapj}

\usepackage{natbib} 
\usepackage{multirow} 
\usepackage[caption=false]{subfig} 
\usepackage{enumerate} 
\usepackage[T1]{fontenc} 
\usepackage{amsmath} 
\usepackage{rotating} 

\def\reff@jnl#1{{\rm#1\/}}

\def\aj{\reff@jnl{AJ}}                  
\def\araa{\reff@jnl{ARA\&A}}            
\def\apj{\reff@jnl{ApJ}}                
\def\apjl{\reff@jnl{ApJ}}               
\def\apjs{\reff@jnl{ApJS}}              
\def\ao{\reff@jnl{Appl.Optics}}         
\def\apss{\reff@jnl{Ap\&SS}}            
\def\aap{\reff@jnl{A\&A}}               
\def\aapr{\reff@jnl{A\&A~Rev.}}         
\def\aaps{\reff@jnl{A\&AS}}             
\def\azh{\reff@jnl{AZh}}                        
\def\baas{\reff@jnl{BAAS}}              
\def\jrasc{\reff@jnl{JRASC}}            
\def\memras{\reff@jnl{MmRAS}}           
\def\mnras{\reff@jnl{MNRAS}}            
\def\pra{\reff@jnl{Phys.Rev.A}}         
\def\prb{\reff@jnl{Phys.Rev.B}}         
\def\prc{\reff@jnl{Phys.Rev.C}}         
\def\prd{\reff@jnl{Phys.Rev.D}}         
\def\prl{\reff@jnl{Phys.Rev.Lett}}      
\def\pasp{\reff@jnl{PASP}}              
\def\pasj{\reff@jnl{PASJ}}              
\def\qjras{\reff@jnl{QJRAS}}            
\def\skytel{\reff@jnl{S\&T}}            
\def\solphys{\reff@jnl{Solar~Phys.}}    
\def\sovast{\reff@jnl{Soviet~Ast.}}     
\def\ssr{\reff@jnl{Space~Sci.Rev.}}     
\def\zap{\reff@jnl{ZAp}}                        
\def\nat{\reff@jnl{Nature}}             

\begin{document}


\title{AMI Observations of the Anomalous Microwave Emission in the Perseus Molecular Cloud}

\author{C.~T.~Tibbs\altaffilmark{1,2}, A.~M.~M.~Scaife\altaffilmark{3}, C.~Dickinson\altaffilmark{2}, R.~Paladini\altaffilmark{4}, R.~D.~Davies\altaffilmark{2}, R.~J.~Davis\altaffilmark{2}, K.~J.~B.~Grainge\altaffilmark{5,6}, R.~A.~Watson\altaffilmark{2}}
\email{ctibbs@ipac.caltech.edu}

\altaffiltext{1}{\textit{Spitzer} Science Center, California Institute of Technology, Pasadena, CA 91125, USA}
\altaffiltext{2}{Jodrell Bank Centre for Astrophysics, School of Physics and Astronomy, The University of Manchester, M13 9PL, UK}
\altaffiltext{3}{Physics and Astronomy, University of Southampton, Highfield, Southampton, S017 1BJ, UK}
\altaffiltext{4}{NASA \textit{Herschel} Science Center, California Institute of Technology, Pasadena, CA 91125, USA}
\altaffiltext{5}{Astrophysics Group, Cavendish Laboratory, J J Thomson Avenue, Cambridge, CB3 0HE, UK}
\altaffiltext{6}{Kavli Institute for Cosmology, Cambridge, Madingley Road, Cambridge, CB3 0HA, UK}

\shorttitle{AMI Observations of the Perseus Cloud}
\shortauthors{Tibbs et al.}


\begin{abstract}

We present observations of the known anomalous microwave emission region, G159.6-18.5, in the Perseus molecular cloud at 16~GHz performed with the Arcminute Microkelvin Imager Small Array. These are the highest angular resolution observations of G159.6-18.5 at microwave wavelengths. By combining these microwave data with infrared observations between 5.8 and 160~$\mu$m from the \textit{Spitzer Space Telescope}, we investigate the existence of a microwave~--~infrared correlation on angular scales of~$\sim$~2~arcmin. We find that the overall correlation appears to increase towards shorter infrared wavelengths, which is consistent with the microwave emission being produced by electric dipole radiation from small, spinning dust grains. We also find that the microwave~--~infrared correlation peaks at 24~$\mu$m~(6.7$\sigma$), suggesting that the microwave emission is originating from a population of stochastically heated small interstellar dust grains rather than polycyclic aromatic hydrocarbons.

\end{abstract}




\section{Introduction}
\label{sec:intro}

A new Galactic emission mechanism, commonly referred to as anomalous microwave emission~(AME), has been observed by a number of experiments aimed at measuring fluctuations in the Cosmic Microwave Background~(CMB) in the frequency range 10~--~100~GHz~\citep[e.g.][]{Leitch:97}. Studies over the last decade have shown that AME is prevalent in a variety of Galactic environments such as molecular clouds~\citep{Watson:05, Casassus:08, Tibbs:10, Planck_Dickinson:11}, dark clouds~\citep{Casassus:06, Ami:09, Scaife:09, Dickinson:10}, H\textsc{ii} regions~\citep{Dickinson:07, Dickinson:09, Tibbs:12}, and in the diffuse interstellar medium~\citep[][]{Leitch:97, Lagache:03, MD:08, Ghosh:12}. AME has also been detected within a star forming region in the external galaxy NGC~6946~\citep{Murphy:10, Scaife:10}. The AME is observed as an excess of emission at cm wavelengths above the other Galactic emission mechanisms and is expected to be a substantial contaminant to the CMB in the frequency range 10~--~100~GHz. Hence, it is of critical importance that we completely comprehend the physical emission mechanism involved. 

The current favored explanation for the source of this emission is electric dipole radiation originating from rapidly rotating very small dust grains~\citep{DaL:98}. Interstellar dust grains are responsible for the mid- to far-infrared~(IR) emission, which is generally thought to be produced by a combination of stochastic and thermal equilibrium emission processes. The stochastic emission observed at mid-IR wavelengths~($\sim$~3~--~70~$\mu$m) is generated by emission from polycyclic aromatic hydrocarbons~(PAHs) and a population of small carbonaceous dust grains, while larger silicate dust grains are responsible for the thermal equilibrium emission at far-IR wavelengths~($\gtrsim$~70~$\mu$m). The spinning dust hypothesis implies that there is an intrinsic link between the AME and the stochastic emission observed in the mid-IR as both are produced by small interstellar dust grains. If the spinning dust mechanism proves to be correct, then AME studies represent a new window in which to probe small interstellar dust grains and the physics involved.


In this analysis we focus on a known region of AME in the vicinity of the Perseus molecular cloud. The dust feature G159.6-18.5 has previously been studied and found to exhibit AME on angular scales of~$\sim$~1~degree~\citep{Watson:05, Planck_Dickinson:11} and~$\sim$~10~arcmin~\citep{Tibbs:10}. Also, a recent study performed by~\citet{Tibbs:11} used IR observations of the region to characterize the dust properties and investigate the physical conditions in which the AME is observed. In this work, we present new observations of the Perseus molecular cloud performed with the Arcminute Microkelvin Imager~(AMI) Small Array~(SA). These observations represent the highest angular resolution observations of this region at microwave wavelengths to date, and allow us to investigate the AME on angular scales of~$\sim$~2~arcmin, providing further insight into the physical processes producing the AME. 

The layout of this paper is as follows. Section~\ref{sec:obs} describes the data used in this analysis and in Section~\ref{sec:corr} we investigate the correlation between the microwave and IR emission. In Section~\ref{sec:discuss} we discuss the results of this investigation and we present our conclusions in Section~\ref{sec:con}.


\section{Observations}
\label{sec:obs}


\subsection{AMI SA Data}
\label{subsec:ami}

The AMI SA is a radio interferometer consisting of ten 3.7~m antennas situated at Mullard Radio Astronomy Observatory, Lord's Bridge, Cambridge. The array has a baseline range of~$\approx$~5~--~20~m and operates in eight  0.75~GHz channels over the frequency range 12~--~18~GHz~\citep{Zwart:08}. However, due to interference from geostationary satellites, the two lowest frequency channels were ignored throughout this analysis.

\begin{deluxetable}{cccc}
\tabletypesize{\normalsize}
\tablecaption{summary of the ami sa observations of the perseus molecular cloud}
\tablewidth{0pt}
\tablehead{
\colhead{Target} & \colhead{RA} & \colhead{Dec} & \colhead{r.m.s. noise} \\
 & \colhead{(J2000)} & \colhead{(J2000)} & \colhead{(mJy~beam$^{-1}$)} \\
 }
 \startdata
Per1  & 03:44:34.6 & +32:08:07.3 & 0.20 \\ 
Per2  & 03:39:57.7 & +31:56:14.1 & 0.26 \\ 
Per3  & 03:37:20.7 & +31:23:29.3 & 0.24 \\ 
Per4  & 03:43:17.6 & +32:00:28.9 & 0.30 \\ 
Per5  & 03:43:02.2 & +31:43:25.7 & 0.32 \\ 
Per6  & 03:42:49.5 & +31:29:34.3 & 0.35 \\ 
Per7  & 03:40:24.5 & +31:15:43.1 & 0.39 \\
\enddata
\label{Table:summary-SA}
\end{deluxetable}

\begin{figure}
\begin{center}
\includegraphics[angle=0,scale=0.46]{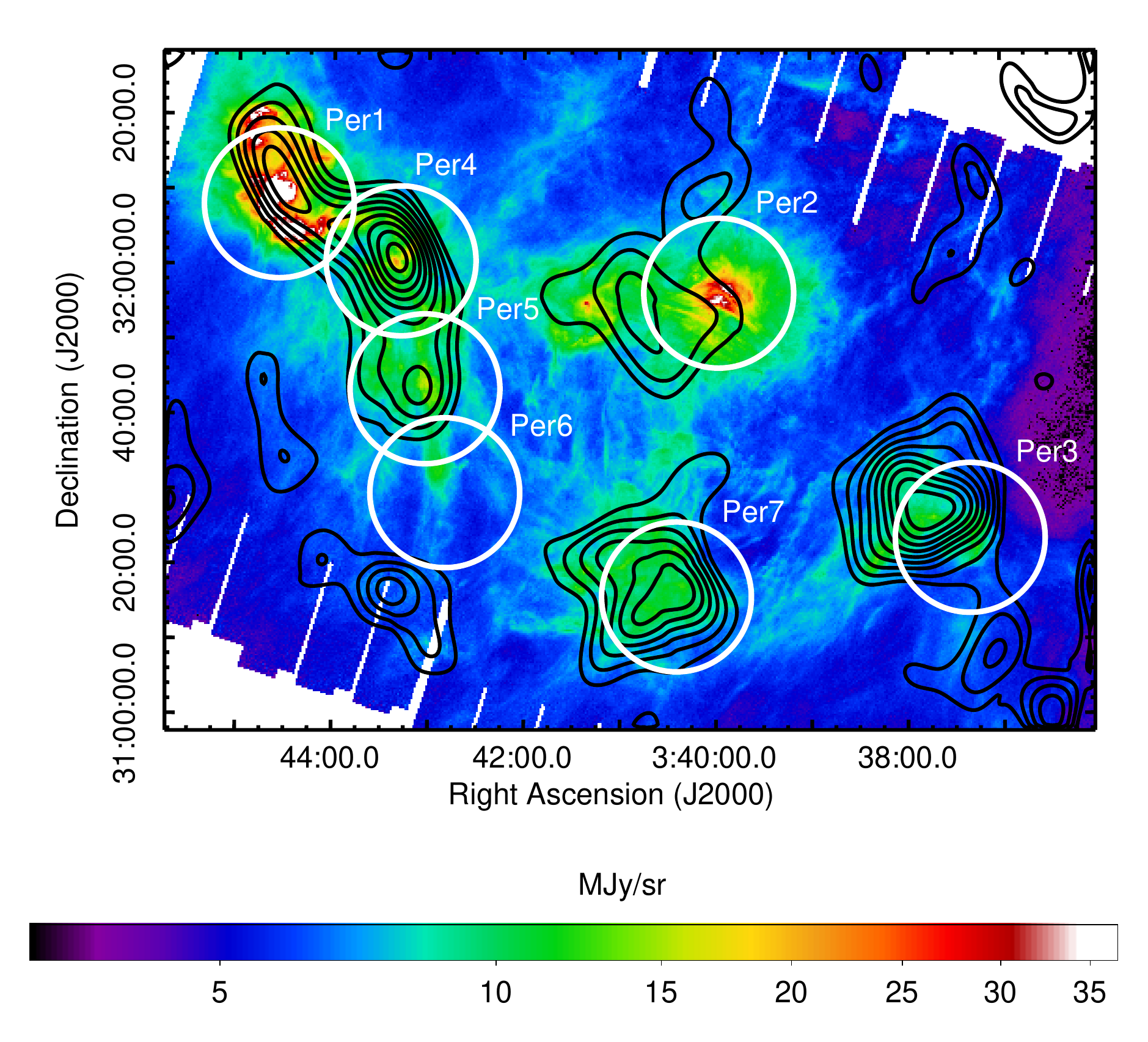} \\ 
\caption{\textit{Spitzer} MIPS 24~$\mu$m map of G159.6-18.5 overlaid with contours from the VSA observations~\citep{Tibbs:10}. Also displayed are the seven AMI SA pointings~(circles represent the~$\approx$~20~arcmin FWHM of the AMI SA primary beam). This illustrates the location of the AMI SA pointings relative to both the AME previously detected and the small scale dust features in this region. The contours correspond to 10, 20, 30, 40, 50, 60, 70, 80, and 90~\% of the peak VSA flux, which is 200~mJy~beam$^{-1}$.}
\label{Fig:VSA-AMI}
\end{center}
\end{figure}

The AMI SA has a typical system temperature of~$\approx$~25~K, and the point source sensitivity of the array is~$\approx$~30~mJy~s$^{1/2}$, which corresponds to~$\approx$~0.2~mJy over an 8~hour observing run. Observations of G159.6-18.5 were performed throughout June and July 2010, with a total of seven individual pointings~(see Table~\ref{Table:summary-SA}). The location of the pointings was chosen to coincide with both the AME features previously observed with the Very Small Array~(VSA) at 33~GHz on angular scales of~$\sim$~10~arcmin~\citep{Tibbs:10}, and the small scale structure present in the mid-IR \textit{Spitzer} maps of the region~\citep{Tibbs:11}. Figure~\ref{Fig:VSA-AMI} displays a map of the 24~$\mu$m emission overlaid with the VSA 33~GHz contours highlighting the strong spatial correlation between the microwave and mid-IR emission on angular scales of~$\sim$~10~arcmin. Also displayed in Figure~\ref{Fig:VSA-AMI} are the seven AMI SA pointings listed in Table~\ref{Table:summary-SA}.

\begin{figure}
\begin{center}
\includegraphics[angle=0,scale=0.46]{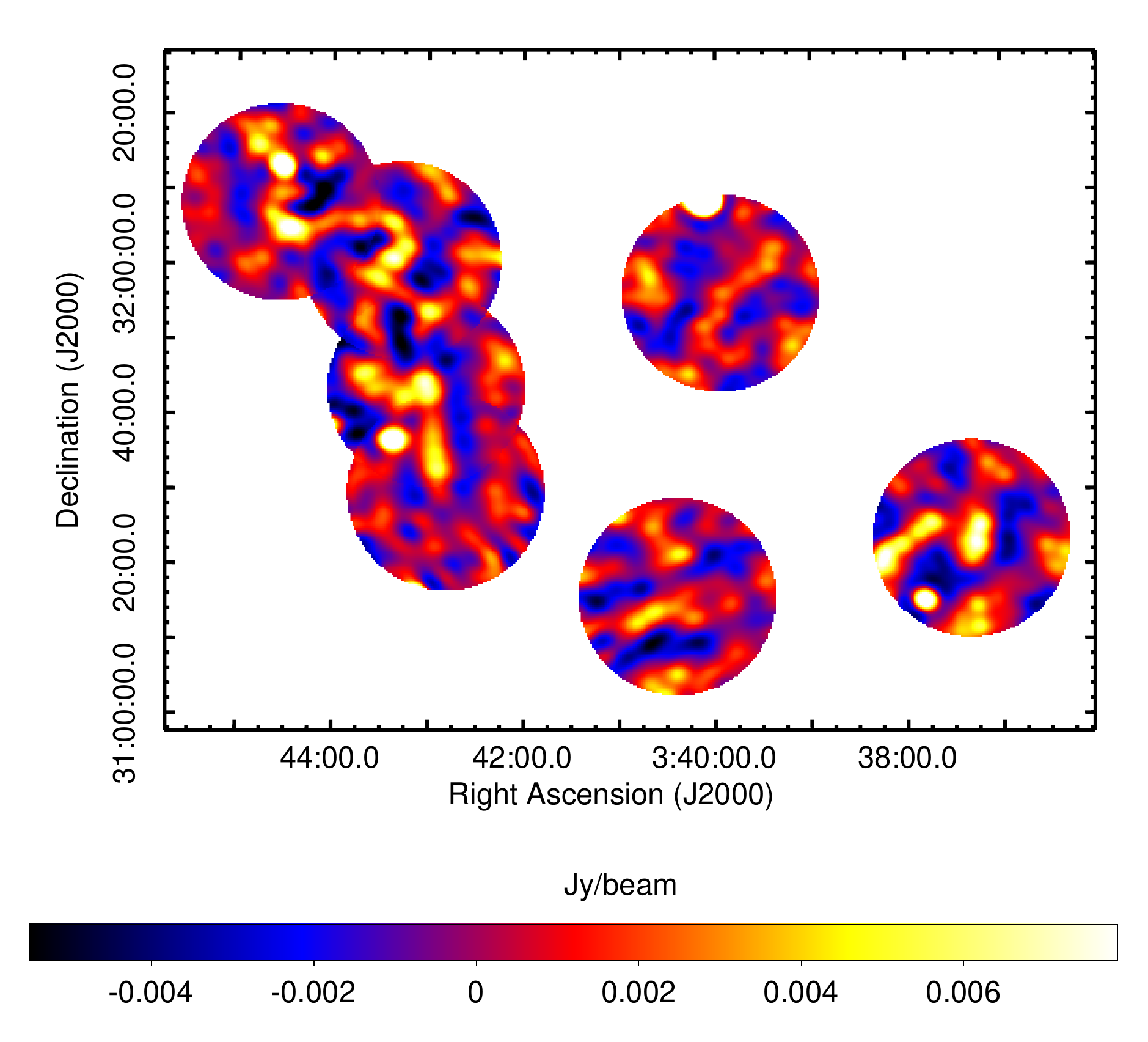} \\ 
\caption{AMI SA mosaic created from the seven pointings of G159.6-18.5. The AMI primary beam FWHM is~$\approx$~20~arcmin and the mean synthesized beam FWHM is 2.6~$\times$~2.1~arcmin$^{2}$ at 16~GHz.}
\label{Fig:AMI_MOS}
\end{center}
\end{figure}

Data reduction was performed using the \textsc{reduce} software package. \textsc{reduce} was specifically designed for AMI observations and automatically flags interference, shadowing, and hardware errors, applies path delay corrections and performs phase and flux density calibrations. Phase calibration was performed using observations of J0359+3220, while flux density calibration was performed using short observations of 3C48. The reduced data were imaged using the \textsc{aips} data package. The task \textsc{imagr} was used to perform both the Fourier Transform and deconvolution for all of the pointings. This task uses a CLEAN based algorithm to perform the deconvolution and we implemented a Briggs robust parameter of 0, which provides a compromise between uniform and natural weighting, and a small loop gain to help retain the extended emission. This resulted in seven CLEAN maps, one for each pointing, and the r.m.s. noise in each of these maps is listed in Table~\ref{Table:summary-SA}. These seven pointings were primary beam corrected and combined to produce a mosaic of the region as displayed in Figure~\ref{Fig:AMI_MOS}. The half power point of the AMI SA primary beam is~$\approx$~20~arcmin at 16~GHz, and the mean AMI SA synthesized beam~(FWHM) for these observations is 2.6~$\times$~2.1~arcmin$^{2}$ at 16~GHz.

\begin{figure*}
\begin{center}
\includegraphics[angle=0,scale=0.68]{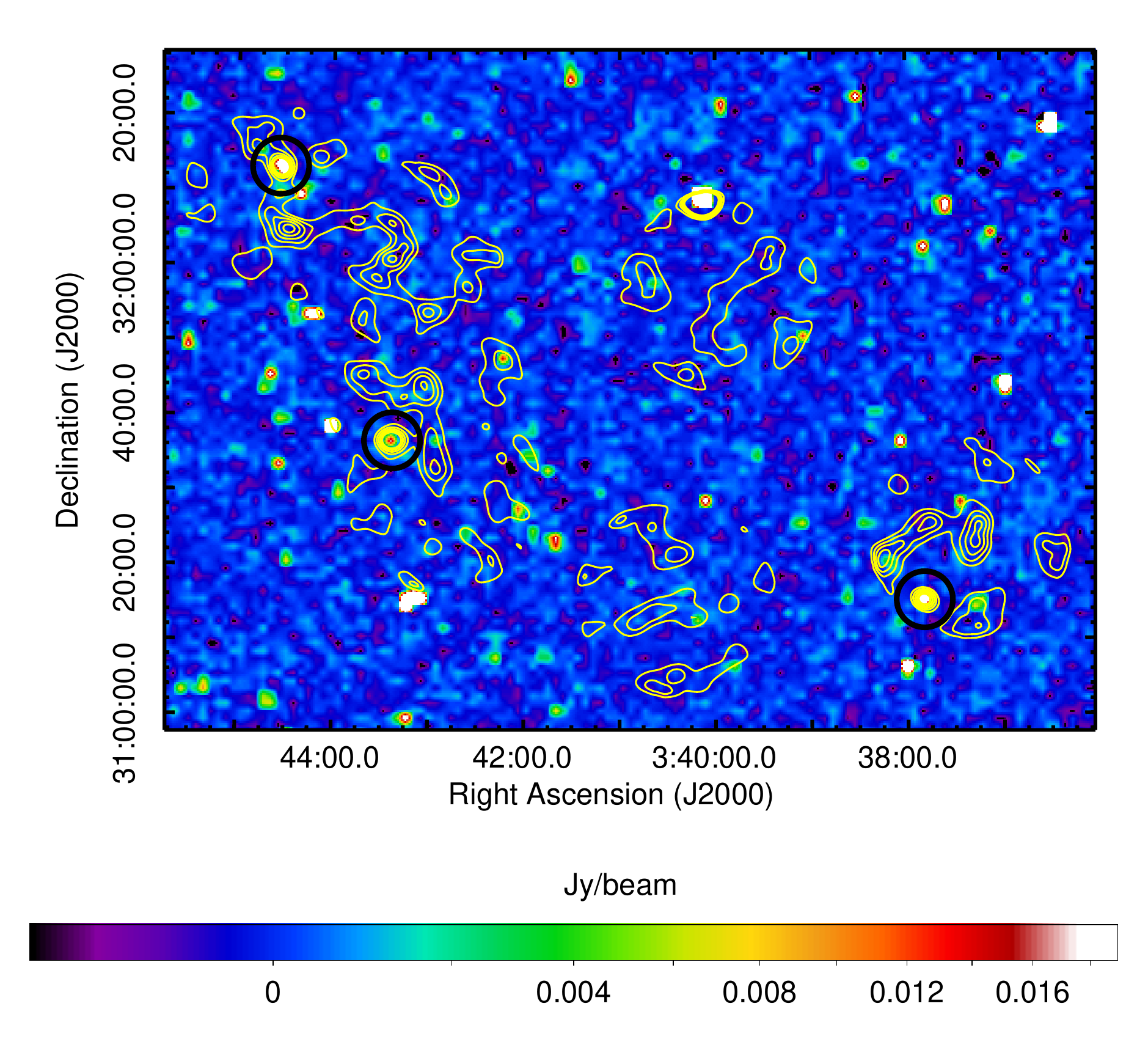} 
\caption{NVSS 1.4~GHz map of G159.6-18.5 overlaid with the AMI SA contours (10, 30, 50, 70, and 90~\% of the peak of the extended emission, which is~$\approx$~10~mJy~beam$^{-1}$). These low frequency radio observations help to identify any compact radio sources in the AMI SA mosaic. We find three sources which are marked with black circles and listed in Table~\ref{Table:radio_sources}. These observations confirm that the emission observed with the AMI SA is not contaminated by point source emission.}
\label{Fig:radio}
\end{center}
\end{figure*}

\subsection{NVSS Data}
\label{subsec:radio}

To identify compact radio sources within the region we used the NVSS 1.4~GHz map~\citep{Condon:98}. Figure~\ref{Fig:radio} displays the NVSS map of G159.6-18.5 overlaid with the AMI SA contours. We find three point sources in the AMI SA mosaic, all of which correspond to a source in the NVSS 1.4~GHz map. These sources, which are highlighted in Figure~\ref{Fig:radio}, are listed in Table~\ref{Table:radio_sources} along with the measured flux density at 16~GHz. The errors include a fitting error combined in quadrature with a 5~\% calibration uncertainty. The location of these sources is such that they do not contaminate the bulk of the extended emission observed in the AMI SA map. It is also apparent that there is a significant lack of correlation between the emission observed with the AMI SA at 16~GHz and the 1.4~GHz emission, which suggests that the source of the emission in the AMI SA mosaic is not originating from either free-free or synchrotron emission. Based on observations performed on both~$\sim$~1~degree and~$\sim$~10~arcmin angular scales, the AME accounts for~$\sim$~80~\% of the cm emission in G159.6-18.5~\citep{Watson:05, Planck_Dickinson:11, Tibbs:10}. Additionally, new observations of G159.6-18.5 obtained with the Robert C. Byrd Green Bank Telescope at 1.4 and 5~GHz~\citep{Tibbs:13} show that the AME observed in this region with the VSA~\citep{Tibbs:10} cannot be accounted for by free-free or synchrotron emission.

\begin{deluxetable}{cccc}
\tabletypesize{\normalsize}
\tablecaption{sources detected in the ami sa mosaic at 16~ghz}
\tablewidth{0pt}
\tablehead{
\colhead{Source} & \colhead{RA} & \colhead{Dec} & \colhead{$S_{16~GHz}$} \\
 & \colhead{(J2000)} & \colhead{(J2000)} & \colhead{(mJy)} \\
 }
 \startdata
NVSS J034433+321255 & 03:44:33.7 & +32:13:04.0 & 18.82~$\pm$~3.30 \\ 
NVSS J033749+311514 & 03:37:49.4 & +31:15:16.4 & 20.43~$\pm$~3.51 \\ 
NVSS J034323+313643 & 03:43:22.7 & +31:36:30.7 & 21.78~$\pm$~3.51 \\ 
\enddata
\label{Table:radio_sources}
\end{deluxetable}


\begin{figure*}
\begin{center}
\includegraphics[angle=0,scale=0.69,viewport=30 30 900 580]{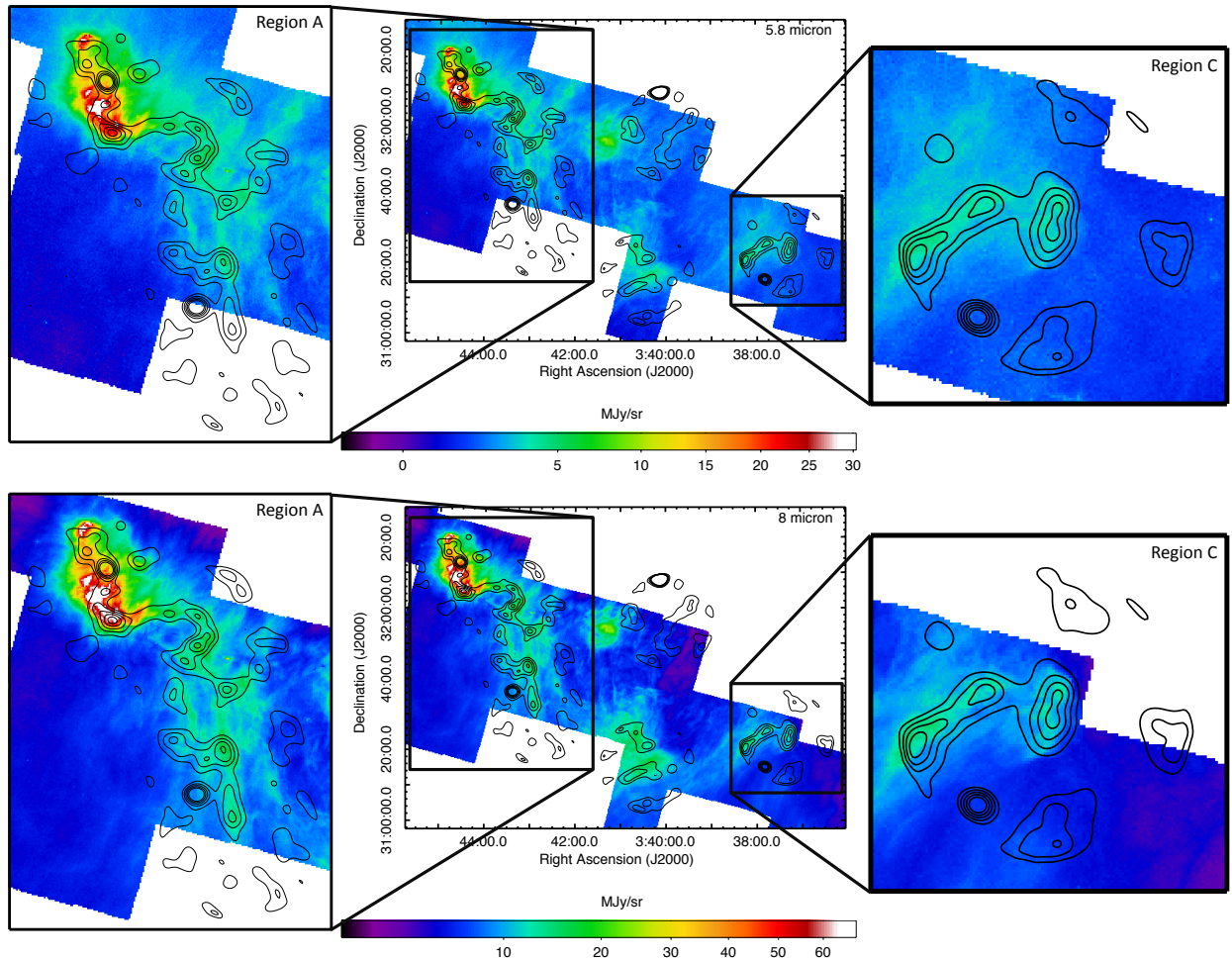} \\ 
\end{center}
\caption{\textit{Spitzer} IRAC maps of G159.6-18.5 at 5.8~(top) and 8~$\mu$m~(bottom) overlaid with the AMI SA contours as displayed in Figure~\ref{Fig:radio}. Also displayed is a zoomed in view of regions A (left) and C (right). The agreement between the emission observed with \textit{Spitzer} and the AMI SA highlight the strong correlation between the microwave and IR emission.}
\label{Fig:Spitzer_Maps_IRAC}
\end{figure*}


\subsection{\textit{Spitzer} Data}
\label{subsec:spitzer}

The Perseus molecular cloud has been observed at mid- to far-IR wavelengths with the \textit{Spitzer Space Telescope} as part of the Cores to Disks~(c2d) Legacy Program~\citep{c2d:03}. However, the IR data used in this analysis are the reprocessed c2d \textit{Spitzer} data that are described by~\citet{Tibbs:11}. These data include IRAC maps of G159.6-18.5 at 5.8 and 8~$\mu$m and MIPS maps at 24, 70, and 160~$\mu$m. These five \textit{Spitzer} maps, all of which have been point source subtracted, provide adequate wavelength coverage incorporating both the stochastically heated dust grains emitting at mid-IR wavelengths and the dust in thermal equilibrium with the exciting radiation field, which emits at longer wavelengths. Figures~\ref{Fig:Spitzer_Maps_IRAC} and~\ref{Fig:Spitzer_Maps_MIPS} display G159.6-18.5 as observed with the two \textit{Spitzer} IRAC bands and the three \textit{Spitzer} MIPS bands, respectively. Contours of the AMI SA emission are overlaid on all five of the \textit{Spitzer} maps.

\vspace{5mm}
\section{Microwave~--~IR Correlation}
\label{sec:corr}

\begin{figure*}
\begin{center}
\includegraphics[angle=0,scale=0.69,viewport=30 30 900 560]{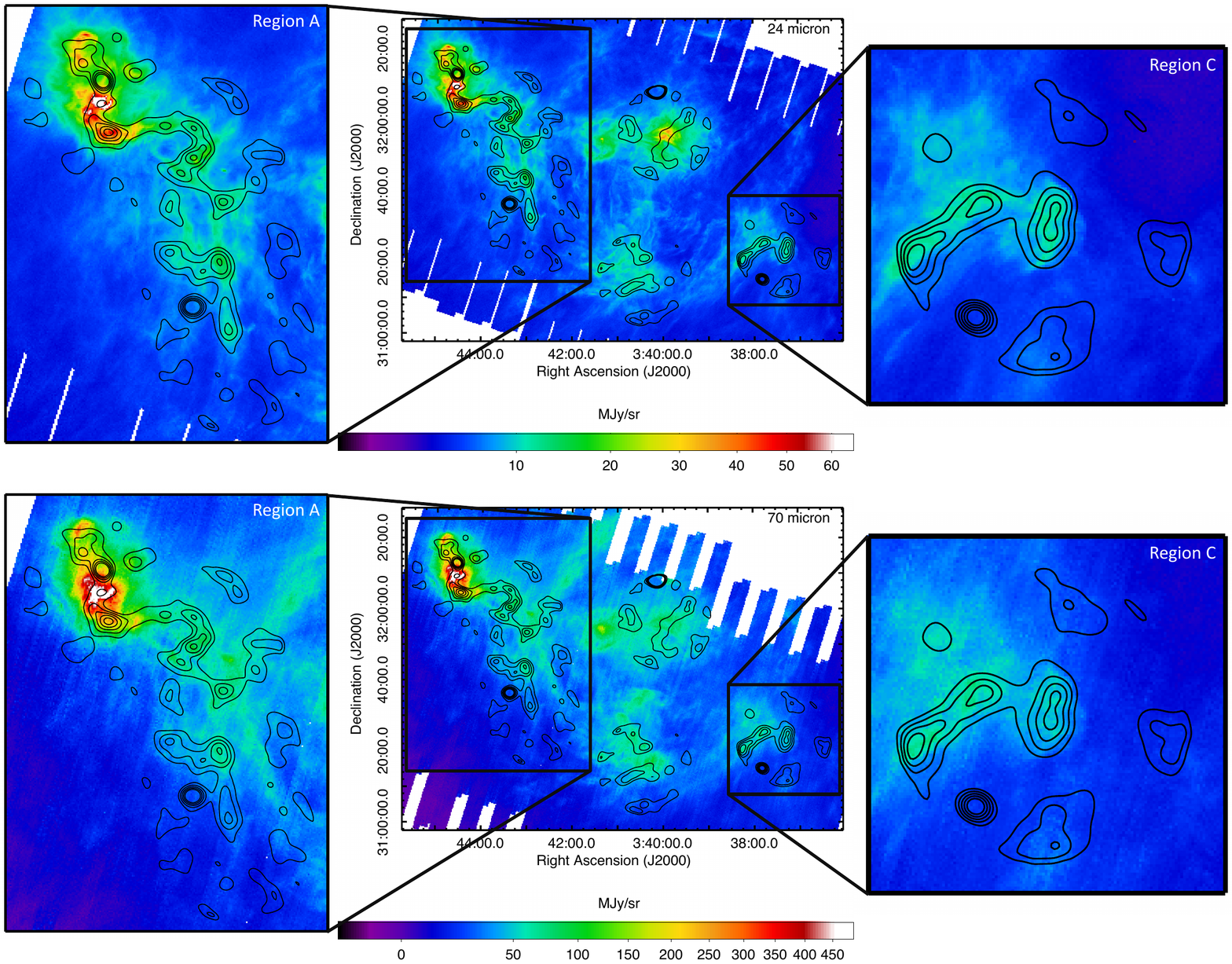} \\ 
\includegraphics[angle=0,scale=0.69,viewport=30 20 900 265]{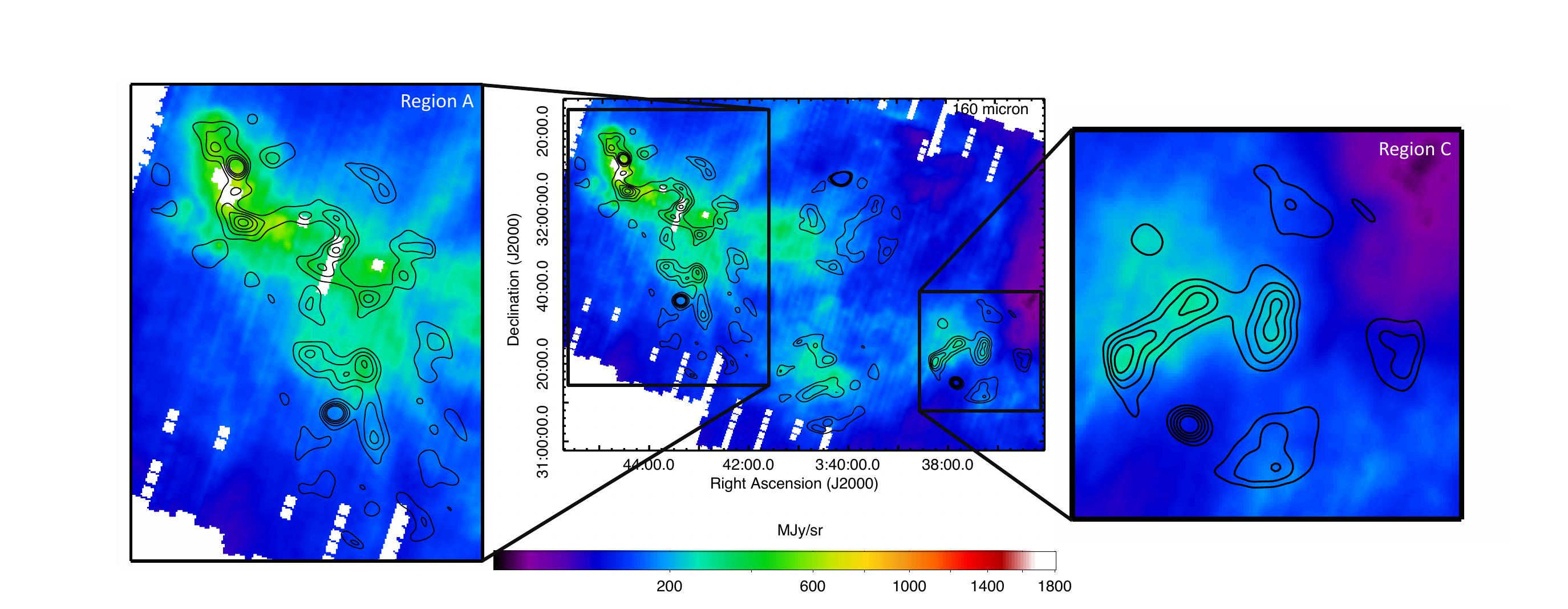} \\ 
\end{center}
\caption{\textit{Spitzer} MIPS maps of G159.6-18.5 at 24~(top), 70 (middle), and 160~$\mu$m~(bottom) overlaid with the AMI SA contours as displayed in Figure~\ref{Fig:radio}. Also displayed is a zoomed in view of regions A (left) and C (right). The agreement between the emission observed with \textit{Spitzer} and the AMI SA highlight the strong correlation between the microwave and IR emission.}
\label{Fig:Spitzer_Maps_MIPS}
\end{figure*}

One of the diagnostics of AME is the strong spatial correlation between the microwave and IR emission. Using the new AMI SA observations presented in Section~\ref{subsec:ami} along with the \textit{Spitzer} data described in Section~\ref{subsec:spitzer}, we investigated this microwave~--~IR correlation in G159.6-18.5 at higher angular resolution than has been performed previously. In Figures~\ref{Fig:Spitzer_Maps_IRAC} and~\ref{Fig:Spitzer_Maps_MIPS} we display G159.6-18.5 with the AMI SA contours overlaid on the \textit{Spitzer} maps at 5.8, 8, 24, 70, and 160~$\mu$m. We defined sub regions within the cloud following a similar naming convention as defined by~\citet{Tibbs:10} for the VSA observations. Region A is defined as the mini mosaic of pointings Per1, Per4, Per5, and Per6, while region B, region C, and region D correspond to pointings Per7, Per3, and Per2, respectively. Looking at Figures~\ref{Fig:Spitzer_Maps_IRAC} and~\ref{Fig:Spitzer_Maps_MIPS}, particularly the zoomed in views of sub regions A and C, it is evident that the microwave emission is spatially correlated with the IR emission. In fact, the microwave emission traces the filamentary structure of the dust, confirming that the microwave~--~IR correlation identified previously in G159.6-18.5 on~$\sim$~10 arcmin angular scales by~\citet{Tibbs:10} is still present on the angular scales observed by the AMI SA~($\sim$~2~arcmin).

\begin{figure*}
\begin{center}$
\begin{array}{cc}
\includegraphics[scale=0.45]{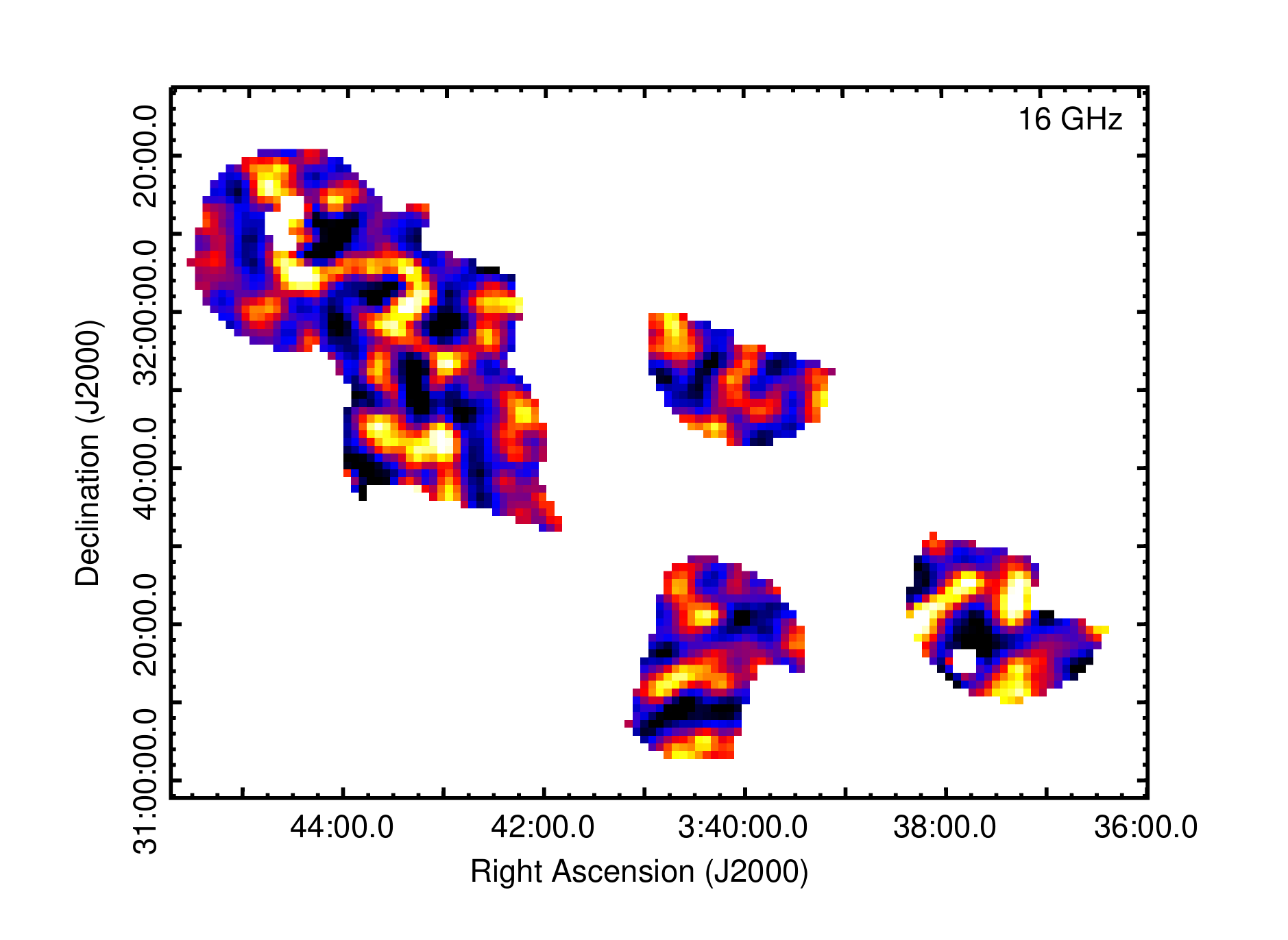} &
\includegraphics[scale=0.45]{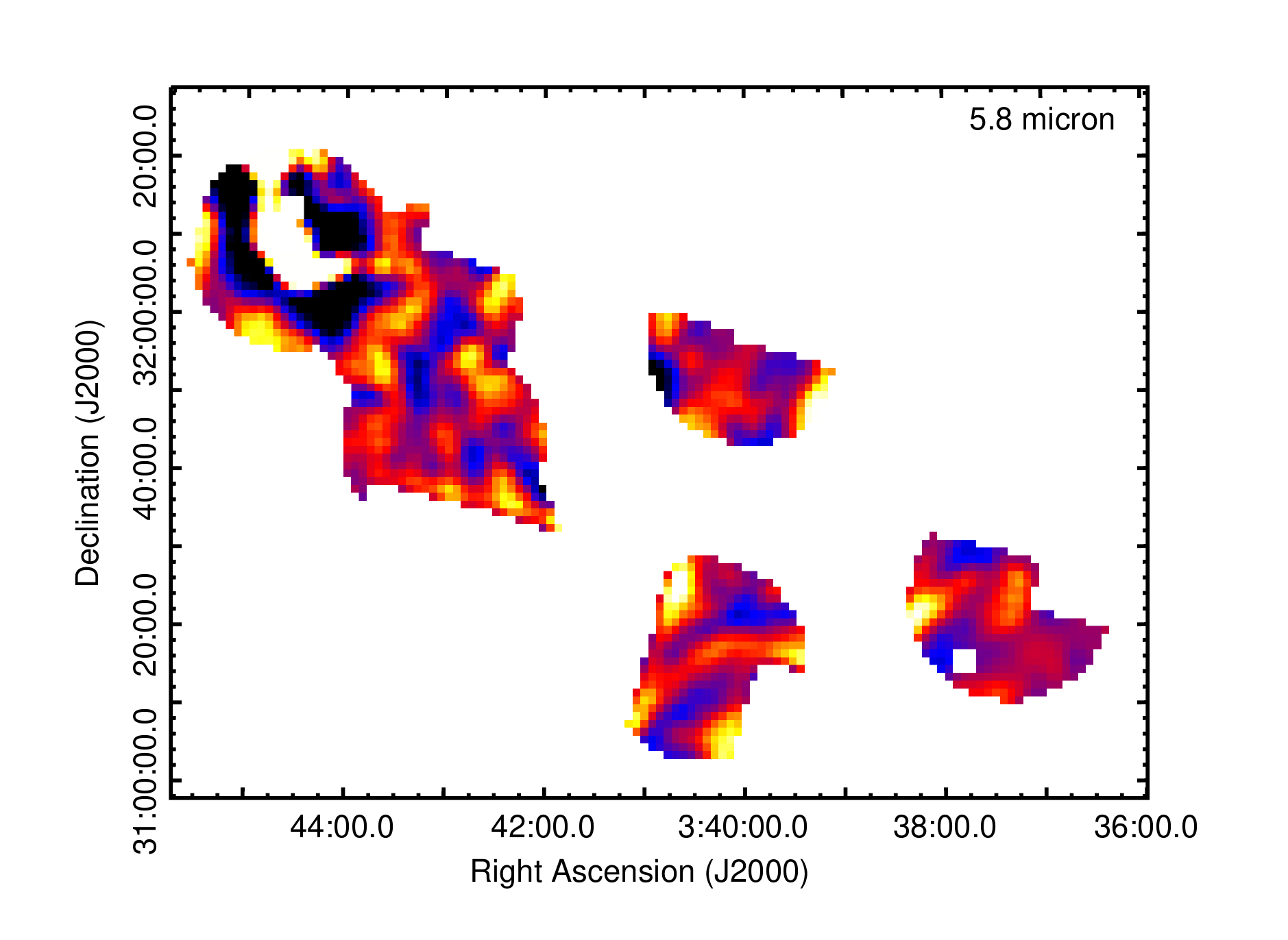} \\
\includegraphics[scale=0.45]{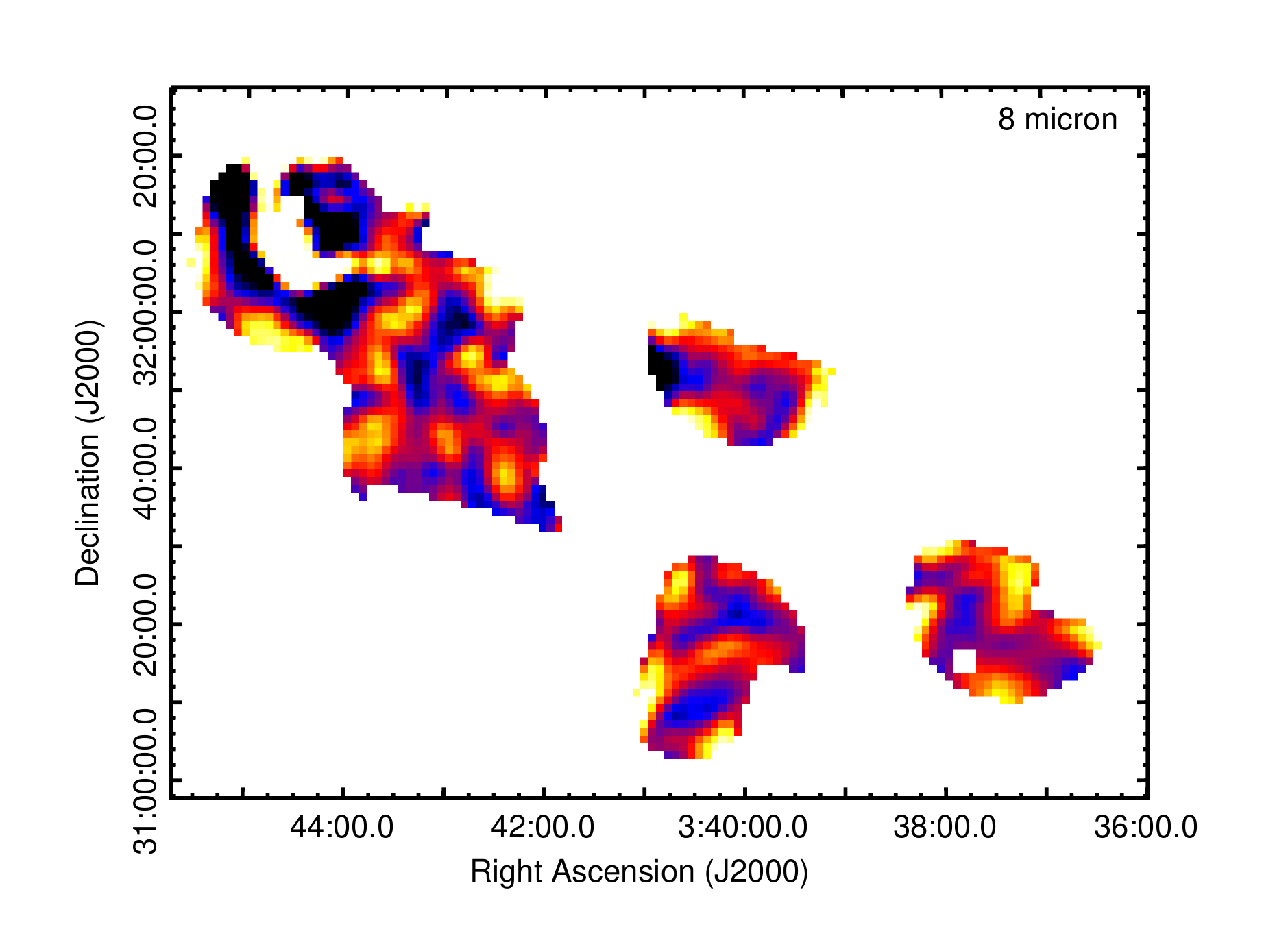} &
\includegraphics[scale=0.45]{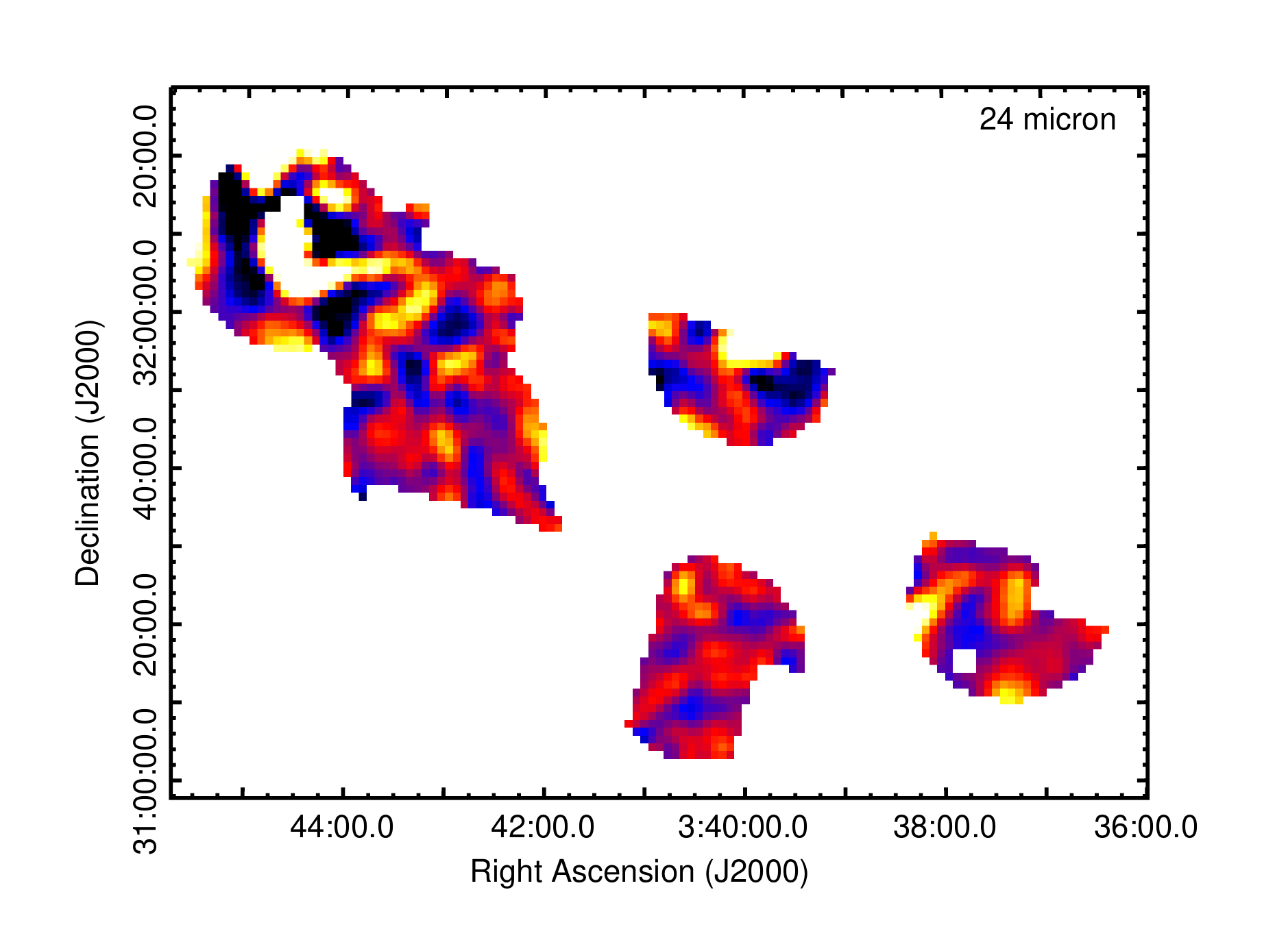} \\
\includegraphics[scale=0.45]{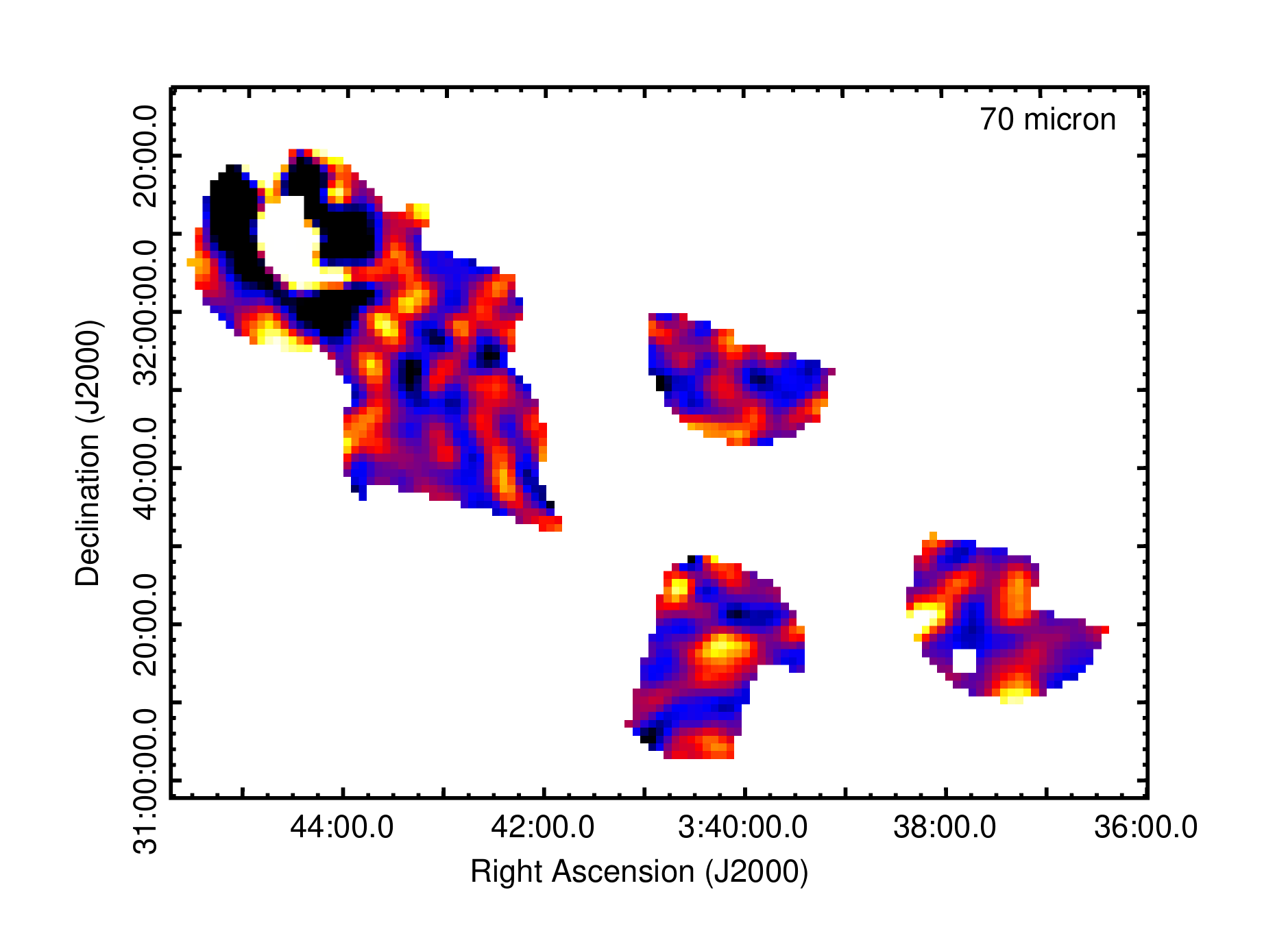} &
\includegraphics[scale=0.45]{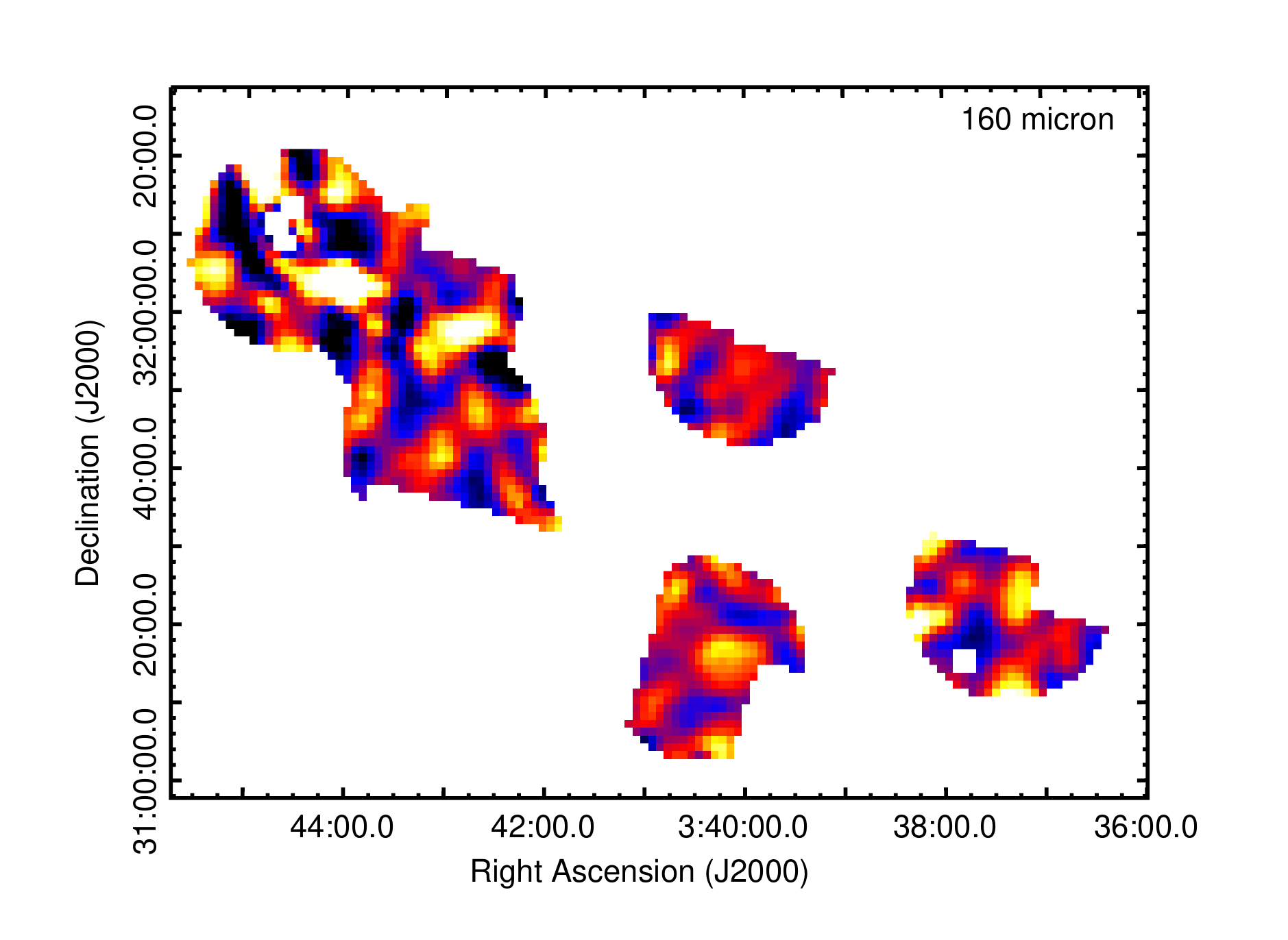} \\
\end{array}$
\caption{The six maps~(AMI SA 16~GHz, \textit{Spitzer} 5.8, 8, 24, 70, and 160~$\mu$m) on which the Pearson correlation analysis was performed. The \textit{Spitzer} maps have been resampled with the AMI SA sampling distribution, and all the maps have been trimmed to match the sky coverage. These maps highlight the strong correlation between the microwave and IR emission.}
\label{Fig:Resampled_Spitzer}
\end{center}
\end{figure*}

To quantify the observed correlation between the microwave and IR emission, we performed a Pearson correlation analysis. However, since the AMI SA is an interferometer most of the extended emission is resolved out, and therefore the AMI SA data and the \textit{Spitzer} data cannot be compared directly. To overcome this issue, the \textit{Spitzer} data must be resampled with the AMI SA sampling distribution to account for the incomplete \textit{u,v} coverage. This task was performed using the \textsc{aips} routine \textsc{uvsub}, which applies the AMI SA sampling distribution to the \textit{Spitzer} data.

After resampling the \textit{Spitzer} data, a Pearson correlation analysis was performed in the image plane between the AMI SA map and the five resampled \textit{Spitzer} maps. Before performing the correlation, all the maps were regridded to a common 1~arcmin pixel grid to ensure the data obeyed the Nyquist sampling theorem, which ensures that each pixel represents approximately independent data points. All six maps~(see Figure~\ref{Fig:Resampled_Spitzer}) were then trimmed to ensure that only the sky coverage common to all the maps was included in the analysis. This also ensures that any artificial edge effects due to the Fourier Transform do not bias the correlation. As we are only interested in the extended emission, the three sources identified in Section~\ref{subsec:radio} were masked and excluded from the rest of this analysis. The results of the correlation analysis are displayed in Figure~\ref{Fig:correlation}, which plots the Pearson correlation coefficient as a function of wavelength. All uncertainties on the correlation coefficients were estimated using the Fisher $r$ to $z$ transformation~\citep{Fisher:15}, and computing the 68~\% confidence interval.

\vspace{5mm}
\section{Discussion}
\label{sec:discuss}

\begin{figure}
\begin{center}
\includegraphics[angle=0,scale=0.50]{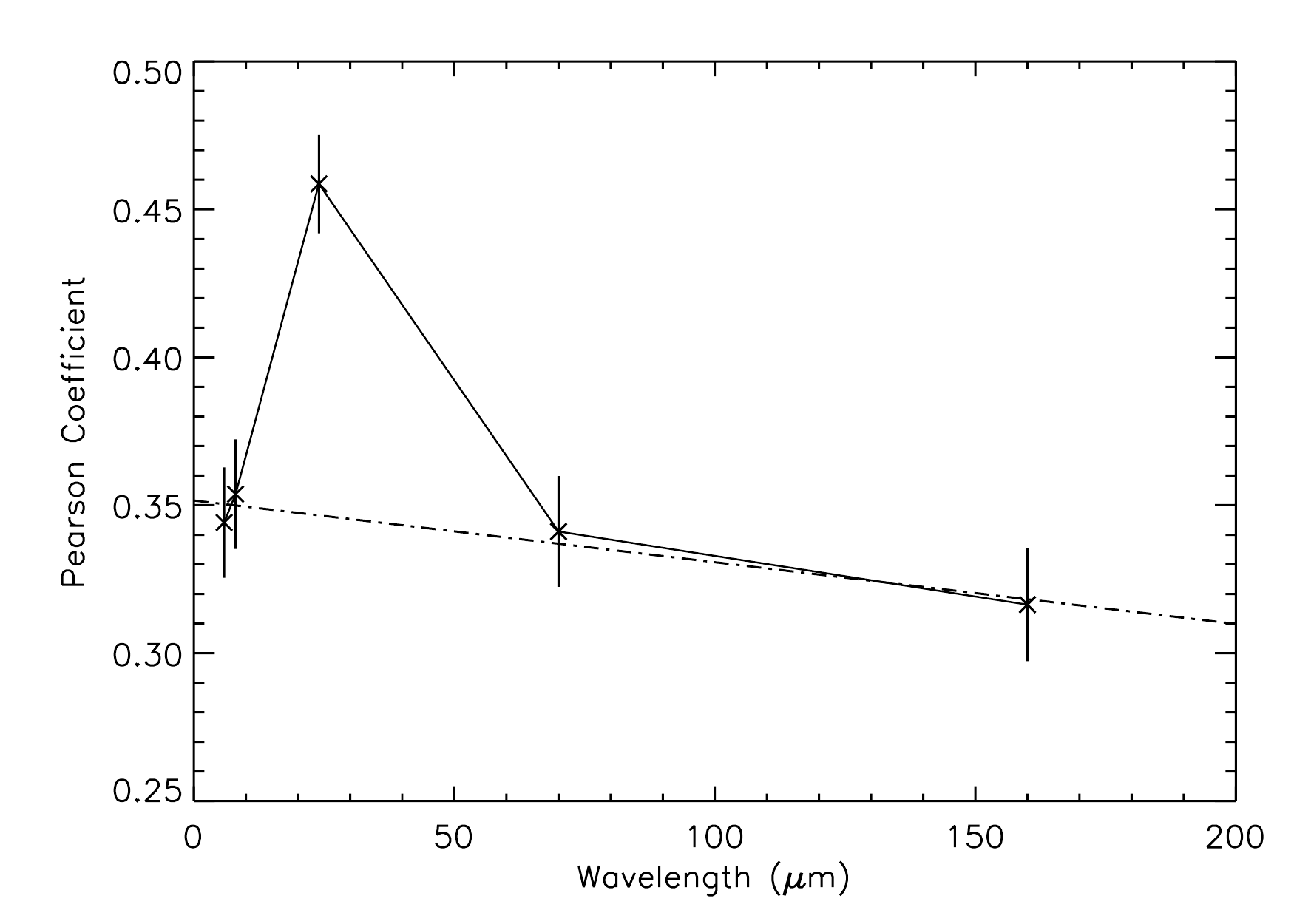} \\
\caption{The computed Pearson correlation coefficient between the emission observed in the AMI SA map and the emission observed in the five resampled \textit{Spitzer} maps for the entire region. Also displayed is a linear fit to the 5.8, 8, 70, and 160~$\mu$m data points~(dot-dashed line). Even ignoring the 24~$\mu$m data point, this fit is non-zero at a level of 1.4$\sigma$, hinting that the correlation is stronger at shorter wavelengths. The 24~$\mu$m data point is in excess of this fit at the level of 6.7$\sigma$, confirming that the correlation peaks at this wavelength. All uncertainties plotted are 1$\sigma$.}
\label{Fig:correlation}
\end{center}
\end{figure}

The results of the Pearson correlation analysis displayed in Figure~\ref{Fig:correlation}, and listed in Table~\ref{Table:Corr_Coef}, show a clear peak in the correlation at 24~$\mu$m at a statistical level of 6.7$\sigma$. In absolute terms, the peak at 24~$\mu$m, with a correlation coefficient of 0.46~$\pm$~0.02, is not a particularly strong correlation, however, as we discuss later, what is interesting is how the relative correlation coefficients vary as a function of wavelength. In addition to the peak in the correlation at 24~$\mu$m, we find that even if we ignore the 24~$\mu$m data point, there is a hint~(1.4$\sigma$) of a non-zero correlation between the Pearson correlation coefficients and the IR wavelengths, which increases at shorter wavelengths. This result tentatively implies that the AME is more correlated with the shorter IR wavelengths, and it is possible to observe this effect by looking at Figure~\ref{Fig:Resampled_Spitzer} where the correlation is clearly stronger with the shorter IR wavelengths. The shorter IR wavelength emission is due to stochastically heated, small dust grains, and this result suggests that the AME is more strongly correlated with the smaller dust grains, which is consistent with the spinning dust hypothesis, where only the smallest dust grains can spin at the frequencies required to produce the observed AME. A similar result was obtained by~\citet{Casassus:06} for LDN1622, where they found that the microwave~--~IR correlation increased towards shorter IR wavelengths.

Ultra-compact H\textsc{ii}~(UCH\textsc{ii}) regions, characterized by densities of~$\gtrsim$~$10^{4}$ cm$^{-3}$, emission measures of~$\gtrsim$~$10^{7}$ pc~cm$^{-6}$, and physical sizes of~$\lesssim$~0.1 pc~\citep{Wood:89}, could possibly contaminate the microwave~--~IR correlation. However, these objects would appear bright and point-like in the AMI SA map, and as shown in Section~\ref{subsec:radio}, there are only three point sources identified in the AMI SA map. These three sources are all extragalactic radio sources, and as stated in Section~\ref{sec:corr}, they were masked for the correlation analysis, thereby ruling out any possible contamination from UCH\textsc{ii} regions. 

\begin{figure}
\begin{center}
\includegraphics[angle=0,scale=0.50]{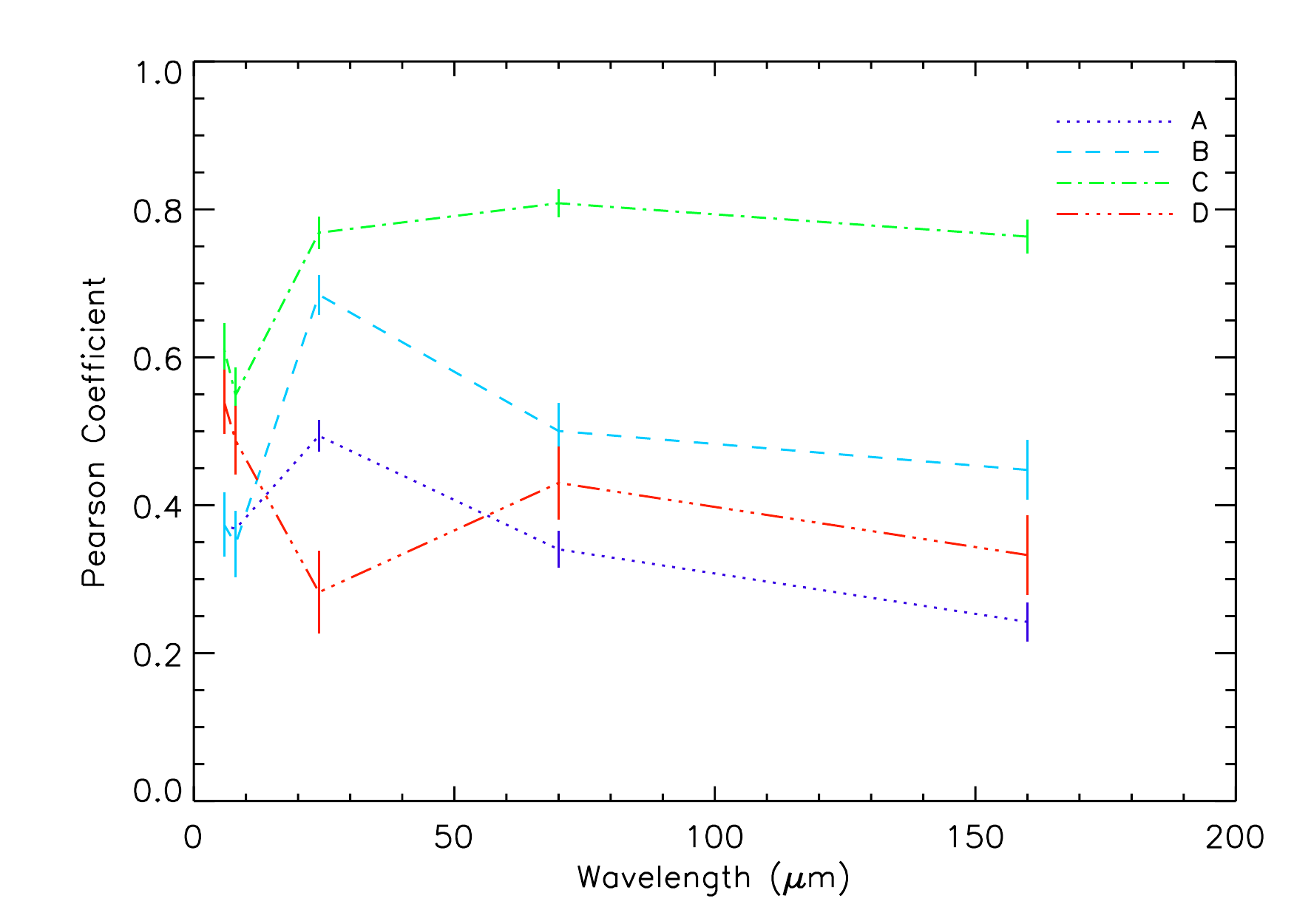} \\
\caption{The computed Pearson correlation coefficient between the emission observed in the AMI SA map and the emission observed in the five resampled \textit{Spitzer} maps for the four sub regions~(A, B, C, and D) within G159.6-18.5. All uncertainties plotted are 1$\sigma$.}
\label{Fig:correlation_subregions}
\end{center}
\end{figure}

To further investigate the microwave~--~IR correlation, we repeated the Pearson correlation analysis. However, this time we performed the analysis separately for each of the sub regions within G159.6-18.5 to allow us to better understand how the microwave~--~IR correlation varies throughout the cloud. The results of the correlation analysis performed on these sub regions are displayed in Figure~\ref{Fig:correlation_subregions}, and tabulated in Table~\ref{Table:Corr_Coef}, and show that there is a clear variation in the microwave~--~IR correlation throughout the entire region. 

For regions A and B, the correlation increases from 160 to 70~$\mu$m and peaks at 24~$\mu$m before falling sharply to 8 and 5.8~$\mu$m. This is similar to what was found for the entire mosaic. Region C shows a flat correlation from 160 to 24~$\mu$m, with a non-significant peak at 70~$\mu$m. This implies that the large dust grains and the small dust grains must be well mixed in this region. However, as is seen in regions A and B, in region C the correlation falls at 8 and 5.8~$\mu$m. For region D, the correlation increases from 160 to 70~$\mu$m,  then decreases at 24~$\mu$m, before rising again at 8 and 5.8~$\mu$m. Region D corresponds to the central region of the shell of G159.6-18.5 containing the B0 V star HD~278942, which is thought to be responsible for the entire shell-like structure. As discussed by~\citet{Tibbs:10, Tibbs:11}, due to the presence of this star, this region is dominated by a very hot dust component that can be seen in the 24~$\mu$m \textit{Spitzer} map. It is therefore likely that the emission at 24~$\mu$m from this region is not due to stochastically heated small dust grains, but is in fact due to thermal emission of large dust grains which have been heated to such temperatures that cause it to emit at 24~$\mu$m. This would therefore explain the lack of correlation between the microwave emission and the 24~$\mu$m emission in this region.

One interesting result is that the correlation does not appear to peak at 5.8 or 8~$\mu$m for any of the sub regions, with the exception of region D, but even this does not show a strong peak. In fact, we find that for regions A, B, and C, the correlation actually decreases from 24~$\mu$m to 8 and 5.8~$\mu$m. These results are consistent with the results found for the entire region, which shows a stronger correlation at 24~$\mu$m than at either 5.8 or 8~$\mu$m. Assuming that contamination from spectral lines is negligible, the IRAC 5.8 and 8~$\mu$m bands, which cover the wavelength of known PAH features at 6.2 and 7.7~$\mu$m, are expected to trace the PAH emission, while the 24~$\mu$m emission is due to stochastically heated small dust grains. The fact that we do not find a peak in the correlation at 5.8 or 8~$\mu$m, but actually at 24~$\mu$m, appears to suggest that the AME is originating from a population of small dust grains rather than PAHs.

\begin{deluxetable*}{cccccc}
\tabletypesize{\normalsize}
\tablecaption{microwave~--~ir correlation coefficients for the entire region and the separate sub regions}
\tablewidth{0pt}
\tablehead{
\colhead{Wavelength} & \colhead{Entire} & \colhead{Region A} & \colhead{Region B} & \colhead{Region C} & \colhead{Region D} \\
\colhead{($\mu$m)} & \colhead{Region} & & & & \\
 }
 \startdata
5.8   & 0.34~$\pm$~0.02 & 0.37~$\pm$~0.02 & 0.37~$\pm$~0.04 & 0.61~$\pm$~0.03 & 0.54~$\pm$~0.04 \\
8      & 0.35~$\pm$~0.02 & 0.37~$\pm$~0.02 & 0.35~$\pm$~0.04 & 0.55~$\pm$~0.04 & 0.49~$\pm$~0.05 \\ 
24    & 0.46~$\pm$~0.02 & 0.49~$\pm$~0.02 & 0.68~$\pm$~0.03 & 0.77~$\pm$~0.02 & 0.28~$\pm$~0.06 \\
70    & 0.34~$\pm$~0.02 & 0.34~$\pm$~0.03 & 0.50~$\pm$~0.04 & 0.81~$\pm$~0.02 & 0.43~$\pm$~0.05 \\
160  & 0.32~$\pm$~0.02 & 0.24~$\pm$~0.03 & 0.45~$\pm$~0.04 & 0.76~$\pm$~0.02 & 0.33~$\pm$~0.05 \\
\enddata
\label{Table:Corr_Coef}
\end{deluxetable*}

A recent study of this region was performed by~\citet{Tibbs:11}, who used IR observations to constrain the properties of the dust and investigated the physical conditions in which the AME is observed. ~\citet{Tibbs:11} found that there was an enhancement in the exciting radiation field in regions of AME. It may be possible that this is what we are observing here, and it is the excitation at 24~$\mu$m, rather than the abundance of the small dust grains, that is producing the correlation with the AME. This is also consistent with the result found in the H\textsc{ii} region RCW175 by~\citet{Tibbs:12}, where the AME was highly correlated with the radiation field. If this is the case, then this suggests that the excitation mechanism of the AME may be more important than the abundance of the carriers. This idea has previously been discussed by~\citet{Casassus:08}, who found that the environmental factors that boost the spinning dust emissivities are more dominant than the abundance of the carriers.


\section{Conclusions}
\label{sec:con}

In order to accurately characterize foreground contaminants to the CMB in the microwave regime, the physical emission mechanism responsible for producing AME must be understood. In this work we present observations of the dust feature G159.6-18.5, located in the Perseus molecular cloud, with the AMI SA. In total, seven pointings were observed and these observations represent the highest angular resolution data of the AME within this region. These seven pointings were compared to the low frequency NVSS data to ensure that there is minimal contamination from free-free emission and emission from UCH\textsc{ii} regions, confirming that the bulk of the AMI SA emission is due to AME. 

We made use of the available high resolution IR observations obtained with the \textit{Spitzer Space Telescope} to investigate the microwave~--~IR correlation. These IR observations were resampled with the AMI SA sampling distribution, to account for the incomplete coverage of the \textit{u,v} plane, and a Pearson correlation analysis was performed. The results of this analysis confirm that the microwave~--~IR correlation observed on angular scales of~$\sim$~10~arcmin is still present on angular scales of~$\sim$~2~arcmin. We find that the microwave~--~IR correlation tentatively increases towards shorter wavelengths, which is what you would expect if the AME is produced by the smallest dust grains. This result is consistent with the spinning dust hypothesis, however, we also found a significant peak in the microwave~--~IR correlation at 24~$\mu$m~(6.7$\sigma$), suggesting that the AME is being produced by small dust grains rather than PAHs. This result is a significant step forward in improving our understanding of the AME, however, this is only one region and only by studying a wide variety of AME regions will we be able to gain a more complete picture of the AME.


\section*{Acknowledgments}

We thank A. Noriega-Crespo and S. Carey for stimulating discussions. We also thank the referee for providing detailed comments that have substantially improved the content of this paper. This work has been performed within the framework of a NASA/ADP ROSES-2009 grant, no. 09-ADP09-0059. CD acknowledges support from an STFC Advanced Fellowship and an EU Marie-Curie IRG grant under the FP7. We thank the staff of the Mullard Radio Astronomy Observatory for their invaluable assistance in the commissioning and operation of AMI, which is supported by Cambridge University and the STFC.



\end{document}